\documentclass[rmp,twocolumn,showpacs,showkeys,preprintnumbers,amsfonts,amsmath,amssymb,a4paper]{revtex4}

\usepackage{graphicx}
\usepackage{dcolumn}
\usepackage{bm}

\begin{document}

\preprint{\emph{Journal Ref: Ukr. J. Phys. 2008. V.53, N 3, p. 261}}

\title{DEPENDENCE OF THE MAGNETIZATION\\ OF AN ENSEMBLE OF
SINGLE-DOMAIN PARTICLES\\ ON~ THE~ MEASUREMENT~ TIME WITHIN
VARIOUS EXPERIMENTAL AND
COMPUTATIONAL METHODS}

\author{A.A. TIMOPHEEV, S.M. RYABCHENKO}

\affiliation{Institute of Physics, National Academy of Sciences of Ukraine\\
46, Nauky Prospect, Kyiv 03028, Ukraine;
}%

\date{\today}

\begin{abstract}
The effect of a measurement time duration on the parameters of
magnetization curves for an ensemble of identical noninteracting
single-domain particles with equally oriented axes under the uniaxial
anisotropy has been specified for different experiment modes, in
particular for the cases of relaxation measurements and the continuous
sweep of a static magnetic field. The relation between a
blocking temperature and experiment characteristics has been
found for these modes. A recursion method to calculate the magnetization reversal curves for such an
ensemble of particles is proposed. By comparing the results of
calculations of the magnetic properties by the recursion and Monte-Carlo methods, an algorithm to establish the relation
between the equivalent measurement time and such parameters of the
Monte-Carlo method as the number of steps and the value of
aperture is suggested.
\end{abstract}

\maketitle

\section{\label{sec:level1}Introduction}

The problems of the magnetism of nanoparticles have attracted the
attention of scientists for many decades. More than
a half-century ago, the English physicists E. Stoner and E.
Wohlfarth developed a simple model of magnetization reversal for
uniaxially anisotropic single-domain particles at $T=0$ K [1].
According to this model, during the reversal, all spins
in each particle turn in such a way that they remain parallel to
each other all the time, i.e. the absolute value of the magnetic
moment for each particle remains constant, and only a mutual
orientation of the magnetic moments of various particles changes.
In such a case, the energy of the ensemble of particles depends only
on one collective variable, for example, on a total magnetization
vector. Within the frame of this model, all the particles
constituting a sample are assumed to have the same shape,
volume $V$, and orientation of the crystallographic
anisotropy axes. The volume fraction $f$ of these single-domain
particles in a specimen is sufficiently small and, thus, the interparticle
interaction can be neglected. The objects, whose behavior
corresponds to this model, are small (in order to satisfy the condition for them to
be single domains) magnetic particles (see, for example, [2])
placed in a nonmagnetic metallic or dielectric matrix.

The density of magnetic energy $U$ for a sample can be represented as the
sum of the energy density of a magnetic anisotropy, which includes the
anisotropy caused by demagnetization fields (a shape-dependent anisotropy
term) and that of the interaction of the magnetic moment with an external
field $H$. In the simplest case of a uniaxial anisotropy with regard for the
first anisotropy constant only, the density of magnetic energy has the form%
\begin{equation}
U={\frac{{f}}{{N}}}{\sum\limits_{i=1}^{N}{U_{i}}}\text{,}  \label{eq1a}
\end{equation}%
where%
\begin{equation}
U_{i}=-K\cos ^{2}(\theta _{i}-\theta _{1})-mH\cos (\theta _{i}).
\label{eq2a}
\end{equation}%
Here, $U_{i}$\ is the density of magnetic energy for the separate $i$-th
particle, $N$ is the number of magnetic particles in the sample volume, $K$
is the first constant of the uniaxial anisotropy of a particle, $m$ is its
magnetization, $\theta _{i}$ is the angle between the magnetic moment of the
$i$-th particle and the magnetic field direction, and $\theta _{1}$\ is the
angle between the easy magnetization axis of a particle (for all the
particles its direction is assumed to be identical) and the magnetic field
direction. For a sample which resides without an external magnetic field as
long as possible, the total magnetization $M$ will turn to zero, since the
numbers of particles with the magnetic moments oriented in parallel and
antiparallel to the easy magnetization axis equal each other. On the
contrary, the value of magnetization becomes finite upon the application of
a magnetic field to the ensemble of particles. Hence, in the case where the
observation duration of such the ensemble is infinitely long, its behavior
will be characteristic of a paramagnet, in spite of the fact that the
particles are ferromagnetic.

Neel [3] and Brown [4] took the fact into consideration that a drastic
removal of the magnetic field from the ensemble of single-domain particles
results in a time decay of the residual magnetization $M(t)$\ according to
the exponential law:%
\begin{equation}
M(t)=M(0)\,\exp \left( {-{\frac{{t}}{{\tau }}}}\right) ,
\label{eq3a}
\end{equation}%
where $M(0)$ is the magnetization value at the initial time moment and $\tau $
is the relaxation time. The latter characterizes the thermally activated
reversal of the direction of the magnetic moment of a separate particle between
two possible minima of its potential energy. The hopping probability obeys
the Arrhenius law which yields
\begin{equation}
\tau ={\frac{{1}}{{\nu _{0}\left( {\exp \left( {-{\frac{{E_{mb}-E_{m1}}}{{kT}%
}}}\right) +\exp \left( {-{\frac{{E_{mb}-E_{m2}}}{{kT}}}}\right) }\right) }}}%
,  \label{eq4a}
\end{equation}%
where $k$ is the Boltzmann constant, $T$ is the temperature, and $E_{m1}$, $%
E_{m2}$, and $E_{mb}$ are the energies for two minima
and a barrier between them, respectively. The expressions for
these energy values contain the product of the particle volume $V$
and the corresponding energy density $U_{i}$ which is dependent
on the relative orientation of the particle magnetic moment and
the field. As a result, the relaxation time $\tau $ strongly depends
on the particle volume, temperature, and the value of the applied
magnetic field.\ If the magnetic field is zero, the energy minima
are equally deep. The application of a finite field makes the
energy of one of them increase, while that of another one
decrease. In a strong magnetic field, when the particle's Zeeman
energy exceeds that of the uniaxial anisotropy, the higher-energy
minimum disappears. For the magnetic particles under discussion,
the typical values of the preexponential factor $\nu _{0}$ in
(\ref{eq4a}) are between $10^{8}$ and $10^{10}$ s$^{-1}$. \ In
the Arrhenius law, this factor is called ``the attempt frequency''.
For estimations, its value can be assumed to equal the precession
frequency for the magnetic moment in an effective magnetic field.
Neel named these materials, which are the independent single-domain
magnetic particles, superparamagnets and called their
quasiparamagnetic behavior as superparamagnetism.

For the superparamagnets, the shape of magnetization curves strongly
varies, depending on the duration $t$ of the measurement process
(measuring time). For each chosen value $t$, a blocking temperature $%
T_{b}$ can be introduced, which divides the whole region of
temperatures into two ones with different magnetization behaviors.
For one of them, the hopping, which occurs during the
measuring time, of the particle magnetic moments between two
energy minima should necessarily be taken into account. But, for the
second one, these effects are not essential and, thus, can be
neglected. It is suitable to choose the temperature, at which the
temperature-dependent relaxation time $\tau $ becomes equal to the
measuring time $t$, as a blocking temperature. For $T\gg T_{b}$, the measuring time
$t\gg \tau $, and the magnetic moment of a particle has enough time
to make multiple jumps between the energy minima. As a result, the
populations of these minima do not differ from the equilibrium
ones and the behavior of such a system of particles will be
close to that of the ensemble of paramagnetic atoms, which is
characterized by the absence of magnetization hysteresis. In this
case, the magnetization of the ensemble of particles can be
described by
formula%
\begin{equation}
M(H)=f\, \overline{m(H)},  \label{eq5a}
\end{equation}%
where the time-averaged particle magnetization, which is identical for all
the particles, equals%
\begin{equation}
\overline {m(H)} = {\frac{{m}}{{{\int\limits_{ - \pi} ^{\pi}
{\exp\left( { - {\frac{{U_{i}  V}}{{kT}}}} \right) }} d\theta
_{i}}} }\quad {\int\limits_{ - \pi} ^{\pi} {\exp\left( { -
{\frac{{U_{i}  V}}{{kT}}}} \right) \cos (\theta _{i} )}} d\theta
_{i} . \label{eq6a}
\end{equation}

The short-time measurements, for which $t<\tau $, correspond to $T<T_{b}$.
In this case, there is no enough time for the transitions between the
energy minima to occur, and the equilibrium populations for the states with
different orientations of the magnetic moments of particles are not achieved
during the measuring time. The system is in a metastable state and the
curves of magnetization reversal display the hysteresis. The coercivity $H_{c}$
depends on the measuring time, anisotropy energy, and temperature. For
the ensemble of uniaxial single-domain particles, Neel and Brown suggested a
simple formula which connects these three parameters:%
\begin{equation}
H_{c}(T)=H_{c}(0){\left\{ {1-{\left[ {{\frac{{kT\ln (t/t_{0})}}{{E_{0}}}}}%
\right] }^{1/\alpha }}\right\} }\text{,}  \label{eq7a}
\end{equation}%
where $t_{0}=\nu _{0}^{-1}$, $E_{0}=KV$ is the height of the energy barrier
between the two minima at $H=0$, and the exponent $\alpha $ is the
parameter which depends on the orientation of a magnetic field relative to
the easy axis of magnetization. For the case of the ensemble of uniaxial
particles, whose easy axes are aligned along the magnetic field, $\alpha =2$%
.. For an arbitrary, but identical for all the particles, orientation of easy
axes (with respect to the magnetic field direction), $\alpha $ is given as
(see [5])%
\begin{equation}
\alpha =0.86+1.14\left( {(\cos ^{2/3}(\theta _{1})+\sin
^{2/3}(\theta _{1}))^{-3/2}}\right) .  \label{eq8a}
\end{equation}%
If the directions of the particles' easy axes are uniformly distributed over
the space, $\alpha =4/3$.

In the literature sources, for the case of $\alpha =2$, one can often find a
representation of formula (\ref{eq7a}) in the form%
\begin{equation}
H_{c}(T)=H_{c}(0)\left( {1-\sqrt{T/{T_{_{b}}^{\text{NB}}}}}\right)
, \label{eq9a}
\end{equation}%
where%
\begin{equation}
T_{b}^{\text{NB}}=\frac{KV}{k\ln (t/t_{0})}  \label{eq10a}
\end{equation}%
is the blocking temperature in the Neel--Brown approximation. It is
worth noting that, depending on the form of the anisotropy energy definition
in the above expressions, the anisotropy constant $K$ can appear with a
factor of 2. In spite of a relative simplicity, formulas (\ref{eq7a}) -- (%
\ref{eq9a}) successfully discribe experimental results.

As was shown in work [6], formulas (\ref{eq7a}) -- (\ref{eq9a})
can be supplemented with the expression accounting for the
dependence of the particle saturation magnetization on
temperature. This will result in a deviation of these expressions
from the power dependence for the temperatures which are too close
to the Curie temperature of ferromagnetic particles. The further
improvements are reduced to the account of the distribution of
particles over sizes or anisotropy values or the account, in
various approximations, of a magnetic dipole-dipole interaction
between the particles.\looseness=1

However, the analytical calculation of the magnetization
hysteresis curves for the systems under consideration meets
serious difficulties even for a minimal number of independent
parameters. At the same time, the power of modern computer
systems makes it possible to carry out such calculations by
numerical methods. One of the difficulties, which arise when one
carries on the numerical calculations and tries to compare their
results with experiment, is related to the correctness of the
identification of a measuring time $t$, which appears in
calculations, with a real duration of experiment. This implies
that a relevant ``protocol'' of measurements should be taken into
account.

In modeling the properties of the ensembles of magnetic
nanoparticles, the method of Monte--Carlo (MC) [7--12] has gained
a significant popularity. However, this method does not include
the `` real'' measuring time. Instead of it, the MC method
contains such parameters as the number of mathematical iterations
(MC steps) and the magnitude of angular aperture used to update
the magnetization direction. At the same time, the literature
sources known to us do not contain an explicit relation between
these parameters of the MC method and the equivalent measuring
time, which corresponds to these calculations. In a few works
(see, for example, [10, 13]), to correlate the number of MC steps
with the measuring time, the results of MC calculations are
compared with the data of actual magnetostatic measurements. The
conclusions of these works are reduced to that the number of MC
steps, being optimal from the viewpoint of the likelihood of the
results obtained and the reasonable duration of calculations,
corresponds to unrealistically short measuring times in real
experiments, even for the calculations carried out with the use of
high-performance modern computers. Though a number of papers
devoted to this problem has been published for the last decade,
the methods how to establish the correspondence between the
equivalent measuring time and MC simulation parameters remain
ambiguous.\looseness=1

It should be noted that the calculation, which would be able to
account for the mode of carrying out the experiment, of the
magnetization curves for the ensemble of single-domain particles
has remained a problem, for a solution of which various approaches
continue to be proposed (see, for example, [14,~15]).

A method developed in this work for the modeling of the magnetization
curves doesn't suffer from the above disadvantage. In what follows, we
call it a recursion method (RM). In the literature sources, we haven't met any
examples of the use of such a method. Its adequacy is grounded on the
favorable outcome of the comparison of its results with those of both the MC
simulations and basic formulas of the Neel--Brown model [3,4] described above.
Basing on such comparison, we will be able to establish a specific relation
between the parameters of MC simulation (the number of MC steps and the
magnitude of angular aperture) and the measuring time which corresponds to
these parameters. The method we offer is suitable for the analysis of the
magnetization curves for superparamagnetic systems consisting of uniaxially
anisotropic particles and comprises the cases where the anisotropy axes are
either parallel to each other or randomly oriented. There are no
restrictions on its utilization to the modeling of the behavior of uniaxial
systems with a nonzero second anisotropy constant and even the systems with
a cubic anisotropy. However, to simplify the analysis, the
modeling is carried out, in what follows, for the ensemble of uniaxial particles, whose axes
are aligned in parallel to the magnetic field direction ($\theta _{1}=0$) with
regard for only the first anisotropy constant.

\section{Dependence of Blocking Temperature on~Measuring Time
for Various Measurement~Protocols}
Consider an ensemble of identical noninteracting spherical single-domain
particles, each of which has volume $V$ and is characterized by the
uniaxial crystallographic anisotropy. We assume that the easy axes of
particles are aligned in parallel to the external magnetic field, i.e. we choose $%
\theta _{1}=0$. We take only the first anisotropy constant into account.
Let us rewrite formula (\ref{eq2a}) in terms of dimensionless units by carrying out
a division of its left and right parts by the anisotropy constant $K$%
\begin{equation}
U_{{\rm red}}=-\cos ^{2}(\theta )-2h\cos (\theta ), \label{eq11a}
\end{equation}%
where $h=Hm/(2K)$ is the dimensionless magnetic field. Here and
below, the index $i$ in the notations of the energy density for a
separate particle and the angle characterizing the direction
of its magnetic moment will be omitted. The parameter $T_{\rm
red}=kT/(KV)$ is used as a dimensionless temperature. Let us also give the
definition of the dimensionless measuring time $t_{\rm red}=t\nu
_{0}$, where $\nu _{0}$ is the parameter which has a frequency
dimension [see expression (\ref{eq4a}) for the probability of the
thermally activated reversal of particle magnetic moments] and
$t$ is a real measuring time in seconds.

For the fields $h\in (-1,1)$, the solutions of the equation
$\frac{\partial U_{\rm  {red}}}{\partial \theta }=0$ give us the
coordinates of two energy minima: $\theta _{m1}=\pi $ and $\theta
_{m1}=0$. The barrier between these minima is observed at $\theta
_{b}=\arccos (-h)$. The reduced energies corresponding to
these angles are $E_{1}=(-1+2h)$, $E_{2}=(-1-2h)$, and
$E_{b}=h^{2}$. Substituting these quantities into formula
(\ref{eq4a}), the expression for the dimensionless relaxation time
$\tau _{\rm red}$, which characterizes the thermally activated
jumps between these minima, can
be written as%
\begin{equation}
\tau _{\rm red}={\frac{{1}}{{\exp }\left[ {-\left( {h-1}\right) ^{2}/T%
}_{\rm red}\right] {+\exp }\left[ {-\left( {1+h}\right) ^{2}/T}_{\rm  {red%
}}\right] }}.  \label{eq12a}
\end{equation}

As was noted above, the blocking temperature $T_{b}$\ depends on the
measuring time $t_{\rm red}$. It is seen that the use of the
condition $t_{\rm red}=\tau _{\rm red}(T_{\rm red}=T_{b}^{r})$ for
the determination of the relaxation time leads to the relation%

\begin{equation}
T_{b}^{r}=1/\ln (2t_{\rm red}).  \label{eq13a}
\end{equation}%
Here,\ $T_{b}^{r}$ is the blocking temperature in a zero magnetic
field taken in the dimensionless form defined above. This
expression doesn't coincide with that for the blocking temperature
in the Neel--Brown approximation, $T_{b^{\ast }}^{r}$\ (a reduction
of $T_{b}^{\rm  {NB}}$ determined from expression (\ref{eq10a})
to a dimensionless unit results in $T_{b^{\ast }}^{r}=1/\ln
(t_{\rm red})$).\ Thus, the Neel--Brown approximation corresponds
to neither the condition $t=\tau (T=T_{b})$ (with expression for
$\tau $ in the form (\ref{eq4a})) nor this relation.

Actually, formula (\ref{eq9a}) along with expression (\ref{eq10a}) for $%
T_{b}^{\rm  {NB}}$ can be obtained from the following
considerations. In
the case of high fields and low temperatures, i.e. when $h/T_{\rm red}\gg 1$%
, we can neglect the second exponential term in (\ref{eq4a}) (or in a
dimensionless expression (\ref{eq12a})). Let us take into account only the
time, which is necessary for the thermally activated reversal of a particle
magnetic moment from a metastable to the basic state, and ignore the
backward jumps. In this case,%
\begin{equation}
\tau _{\rm red}^{\ast }\approx \left\{ \exp \left[ -(h-1)^{2}/T_{\rm  {red%
}}\right] \right\} ^{-1}. \label{eq14a}
\end{equation}

A formal extrapolation, which is not strictly accurate, of this expression
to the zero magnetic field and its equating with a measuring time leads
to a definition of the effective blocking temperature $T_{b^{\ast }}^{r}$
in this approximation as%
\begin{equation}
T_{b^{\ast }}^{r}=1/\ln (t_{\rm red}),  \label{eq15a}
\end{equation}%
i.e. to the formula which was obtained by means of reduction of (\ref{eq10a}%
) to the dimensionless units.\ It is seen that $T_{b}^{r}$ (see (\ref{eq13a}%
)) and $T_{b^{\ast }}^{r}$ (see (\ref{eq15a})) are connected by a simple
relation $(T_{b^{\ast }}^{r})^{-1}=$ $(T_{b}^{r})^{-1}-\ln (2)$.

The most important point in the approximation [3,4] is probably
the expression for the temperature dependence of coercivity [see
(\ref{eq7a}) and (\ref{eq9a})]. At low temperatures, the criterion
$h/T_{\rm red}\gg 1$ can already be fulfilled for a coercive field
($h=h_{c}$) and, thus, approximation (\ref{eq14a}), along with
the definition, which follows from it, of the blocking
temperature, i.e. $T_{b^{\ast }}^{r}$ becomes justified.

Consider the question as to which type of experimental data
and measurement mode corresponds the Neel--Brown approximation in more details. Expression (%
\ref{eq9a}) with definition (\ref{eq10a}) [or (\ref{eq12a})\
in dimensionless units] for a blocking temperature can be obtained
proceeding from the assumption that the field which corresponds to
coercivity is the one, for which $\tau _{\rm red}(h=h_{c})=t_{\rm
red}$. Then the
relation%
\begin{equation}
t_{\rm red}\;\approx \left\{ \exp \left[ {-{\frac{{\left( {h_{c}(T_{%
\rm  {red}})-1}\right) ^{2}}}{{T_{\rm red}}}}}\right] \right\} ^{-1}
\label{eq16a}
\end{equation}%
will be valid for the low temperature region of the coercivity vs
temperature dependence. After taking the logarithm on both left and right
parts of (\ref{eq16a}), we obtain the expression for the temperature
dependence of coercivity:%
\begin{equation}
h_{c}(T_{\rm red})=\;1-\sqrt{T_{\rm red}\ln \,(t_{\rm
red})}.\label{eq17a}
\end{equation}

Defining a blocking temperature as that, at which the extrapolation of a low
temperature part of the temperature dependence of coercivity with either
formula (\ref{eq9a}) or (\ref{eq17a}) reaches zero value for a preset
measuring time, we can write the expression for the temperature
dependence of coercivity as%
\begin{equation}
h_{c}(T_{\rm red})=1-\sqrt{T_{\rm red}/T_{b^{\ast }}^{r}},
\label{eq18a}
\end{equation}%
where $T_{b^{\ast }}^{r}$ is determined from formula (\ref{eq15a}).

However, one should keep in mind that, to derive Eq. (\ref{eq18a}), we had to
use assumption (\ref{eq16a}) which is not completely accurate in the
strict sense. Moreover, it should be remembered that, to obtain (\ref{eq15a}%
), we used one more approximation which consisted in the
extrapolation of the low-temperature high-field part of the $\tau
_{\rm red}^{\ast }$ vs $h $ dependence to the zero field. In fact,
the measuring time, which goes into these equations, can be
interpreted as the time, during which the system relaxes from the
magnetosaturated state after the instantaneous switching-on of the given field. Such a definition
of the measuring time is related to the relaxation measurements.
For the relaxation experiments under consideration, the time
dependence of
the magnetization is given as%
\[ M(h,t_{\rm red} ) = M_{\rm equ} (h) + (M_{0} {\rm sign}(h_{\rm
satur} ) - \]
\begin{equation}
\label{eq19a} -M_{\rm equ} (h))\exp \left( - {\frac{{t_{\rm red}}}
{{\tau (h,T_{\rm red} )}}}\right) .
\end{equation}
Here, $M_{\rm  {equ}}(h)$ is the equilibrium   magnetization in a
field $h$,
which for the system under study is determined as $M_{\rm  {equ}%
}(h)=M_{0}\tanh (2h/T_{\rm red})$, sign$(h_{\rm  {satur}})$ is the
sign of the saturation field, and $M_{0}$ is the saturation
magnetization. At the
same time, the coercivity is determined from the condition $M(h_{c},t_{\rm  {%
red}})=0$ which differs from the condition $t_{\rm red}=\tau _{\rm  {red}%
}(h_{c},T_{\rm red})$. The latter, perhaps, might be used as an
approximate
condition. At low temperatures and sufficiently high coercivities (when $%
h_{c}/T_{\rm red}\gg 1$, which is a criterion of the applicability of
approximation (\ref{eq14a})), it is either a relation $\exp \left[ -t_{\rm red}/\tau _{%
\rm  {red}}(h_{c},T_{\rm red})\right] \approx 1/2$ or $t_{\rm red}=\tau _{%
\rm  {red}}(h_{c},T_{\rm red})\ln (2)$ that would better satisfy the
requirement $M(h_{c},t_{\rm red})=0$. Thus, we will use them instead of (\ref%
{eq16a}). The utilization of such an approximation makes it possible to obtain
the equation for the temperature dependence of coercivity. This equation has the
same form as formula (\ref{eq9a}), in which the effective blocking
temperature $T_{b^{\ast }}^{r}$ is substituted by $T_{b^{\ast \ast }}^{r}$.
The latter coincides with neither (\ref{eq10a}) nor (\ref{eq15a}); it equals%
\begin{equation}
T_{b^{\ast \ast }}^{r}=\;{\frac{{1}}{{\ln \,[t_{\rm red}/\ln (2)]}}}={%
\frac{{1}}{{\ln (t_{\rm red})-\ln [\ln (2)]}}}\rm  {.}
\label{eq20a}
\end{equation}%
It is seen that $(T_{b\ast \ast }^{r})^{-1}=$ $(T_{b\ast }^{r})^{-1}-{\ln
[\ln (2)]=}T_{b}^{r}-{\ln (2)-\ln [\ln (2)]}$.

To correctly determine $T_{b}$\ and obtain the expression for the coercivity
vs $T$ dependence, one should not only find the measuring time, but also
take the measurement protocol for the magnetic characteristics of a
system of particles into account.

Consider the relaxation experiments where the saturating magnetic field, which
acts on the system, is instantaneously (or rapidly enough in real conditions so that
$t\ll \tau $) transformed into some other field $h$,
after which the system relaxes with
time. In this case, let the measurements of the magnetization be continuously carried out during a time interval $t_{%
\rm  {red}}$. Consider the ways of the determination of both
$T_{b}^{r}$ [see (\ref{eq13a})] and the effective blocking
temperature $T_{b^{\ast }}^{r}$ [see (\ref{eq18a})] to make it
possible to specify the temperature dependence of coercivity. It
is pertinent to take a temperature at which $\tau _{\rm
red}(h=0)=t_{\rm red}$ [with regard for (\ref{eq12a}) for $\tau
_{\rm red}$] as $T_{b}^{r}$. If this condition is fulfilled, the
zero-field magnetization (the so-called remanent magnetization)
has to be $e$ times less than the saturation magnetization (here
and below, the number $e$ means the base of the natural
logarithm). To measure the coercivity according to this method and
to determine the blocking temperature from the $h_{c}(T_{\rm
red})$ dependences, it is necessary at each measurement
temperature and at a certain field which has the opposite sign
relative to the initial saturation field, to record a time point,
at which the magnetic relaxation curve crosses the zero value. The
corresponding field will equal the coercivity at the given values
of temperature and measuring time. At low temperatures, it will
correspond to the solution of the
equation%
\[ t_{\rm red} =  [\exp\,\left( { - {\frac{{\left( {1 - h_{c}
(T_{\rm red} ,t_{\rm red} )} \right)^{2}}}{{T_{\rm red}}} }}
\right) +\]

\begin{figure}
\includegraphics[width = 3.2 in]{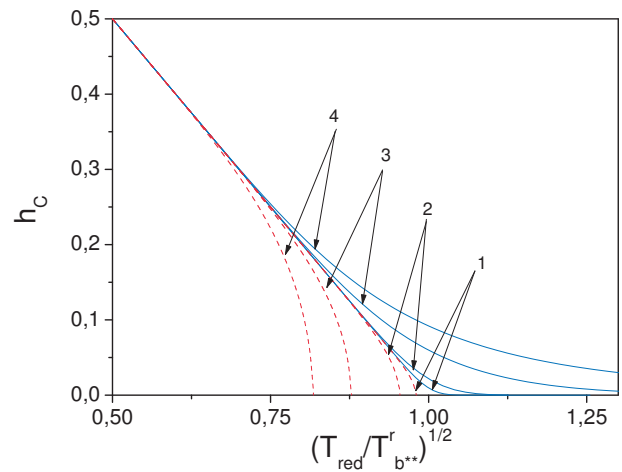}
\caption{\label{fig:epsart} Temperature dependences
of coercivity calculated for different measuring times: $t_{\rm
red}=4$, $10$, $10^{3}$, and $10^{7}$ for the pairs of curves
\textit{1}, \textit{2}, \textit{3}, and \textit{4},
respectively. Dashed and solid lines represent the solutions of Eq. (\ref%
{eq21a}) and $M(h_{c},t_{\rm red})=0$, respectively.}
\end{figure}



\begin{equation}
\label{eq21a} +\exp\,\left( { - {\frac{{\left( {1 + h_{c} (T_{\rm
red} ,t_{\rm red} )} \right)^{2}}}{{T_{\rm red}}} }} \right)]^{ -
1}  \ln 2 .
\end{equation}
Then, for each value of\ $t_{\rm red}$, one should plot the dependence $%
h_{c}(T_{\rm red})|_{t_{\rm red}}$ and find $T_{b\ast \ast }^{r}(t_{%
\rm  {red}})$ from the approximation of its low temperature part
by
formula (\ref{eq18a}), in which $T_{b\ast \ast }^{r}$ substitutes for $%
T_{b\ast }^{r}$. A solution of Eq. (\ref{eq21a}), which describes the
temperature dependence of coercivity, can be found only numerically. As was
noted above, it gives a correct criterion for the determination of $h_{c}(T_{%
\rm  {red}},t_{\rm red})$ only for the temperatures lower than
$T_{b\ast \ast }^{r}$. To find the temperature dependence of
$h_{c}$ over the whole
range of temperatures, it is necessary to solve the equation $M(h_{c},t_{%
\rm  {red}})=0$ with the use of expressions (\ref{eq12a}) and
(\ref{eq19a}) for $\tau _{\rm red}$ and
$M(h,t_{\rm red})$, respectively. It should be noted that the
$h_{c}(T_{\rm red})$\ dependence found in such a way doesn't
exhibit a sharp break by turning into zero at the blocking
temperature. On the contrary, it diminishes smoothly over a
certain temperature range above the blocking temperature.

Figure 1 shows the temperature dependences of coercivity for various
measuring times $t_{\rm red}$ calculated by means of the
numerical solution of Eq. (\ref{eq21a}) and the equation $M(h_{c},t_{\rm
red})=0$ with the components described above. The quantities
$\sqrt{T_{\rm red}/T_{b\ast
\ast }^{r}(t_{\rm red})}=\sqrt{T_{\rm red}\cdot \left\{ \ln (t_{\rm  {%
red}})-\ln \left[ \ln (2)\right] \right\} }$ are taken as the units of the
abscissa axis.

As is seen from the figure, all the curves coincide with each
other at low temperatures. Extrapolation of the low temperature
branches to higher $x$ values gives $\sqrt{T_{\rm red}/T_{b\ast
\ast }^{r}(t_{\rm red})}=1$. However, the real curves do not
follow the extrapolated one. On the contrary, they diverge to
different sides in the vicinity of the effective blocking
temperature. The curves obtained from Eq. (\ref{eq21a}) turn
into
zero below $T_{b\ast \ast }^{r}$, while those obtained from the condition $%
M(h_{c},t_{\rm red})=0$ diminish smoothly above $T_{b\ast \ast
}^{r}$. For very great values of $t_{\rm red}$ ($t_{\rm
red}>10^{5}$), the curves remain linear practically to $T_{b\ast
\ast }^{r}$. It should be noted that, taking
$\nu _{0}\approx 10^{8}\div $ $10^{10}$ s$^{-1}$ into account, only such values
of $t_{\rm red}$ are characteristic of real measurements. The
shorter the measuring time, the lower is the temperature, at which the
curves start to deviate from the linear law which corresponds
to formula (\ref{eq18a}) where the effective blocking temperature equals $%
T_{b\ast \ast }^{r}$. At the same time, both the temperature smearing of a
sharp transition in the vicinity of $h_{c}=0$ and the transition retention
to the temperatures above the blocking temperature disguises a deviation of
the blocking temperature from $T_{b\ast \ast }^{r}$ at short (almost
unlikely in practice) measuring times. For this reason, if one employs
the relaxation method to measure a coercivity, the deviation from
formula (\ref{eq18a}), in which $T_{b\ast \ast }^{r}$ serves as the
effective blocking temperature, will appear only in the form of the
aforementioned smearing. The temperature smearing of the transition, which
is observed as $h_{c}$ tends to zero, seems to be natural, since the
hysteresis is a manifestation of the metastability at a finite measuring
time. Such a hysteresis will occur, to a greater or lesser extent, at any
finite temperatures. It is worth noting that the coercivity, even being
negligibly small at $T>T_{b}$, doesn't turn into zero.

In principle, the relaxation measurement protocol considered above is
used in practice. However, the protocol of the continuous sweep of a magnetic
field (CSMF) with certain rate is more often used for magnetostatic measurements. In the
course of its implementation, the relaxation of the magnetization to its
equilibrium value occurs in a magnetic field, which continuously changes.
The case where the sweeping rate is infinitely small corresponds to the
infinitely great measuring time. In this case, the system is in an
equilibrium state, and such a case corresponds to $T_{b}\rightarrow 0$.
For the regions where superparamagntism becomes clearly apparent and $%
T_{b}\neq 0$, the sweeping rate becomes comparable with the relaxation time $%
\tau $. Under these conditions, the concept of a measuring time should
be made more specific and related to the experiment conditions and the
blocking temperature definition.

In the case of the CSMF studies, the simplest and natural way to
analyze and describe the system properties may be the analysis of
a hypothetical protocol of measurements, in which the whole range of
the field sweep is divided into equal intervals. In this case,
the magnetic field sweep can be regarded as a series of jumps,
each of which being characterized by a specific waiting time
$t_{w}$ after the previous jump. In this case, the magnetization
for each field point will relax during $t_{w}$ from the value at
the previous point. Such a method can be called as recursive, since, in order
to describe the magnetization relaxation at the $n$-th field
point, one should reconstruct a successive series of magnetization
relaxations for all $n-1$ previous points. In the limit where
the interval between successive points tends to zero, we obtain
the CSMF protocol. It is appropriate to define the measuring time
as a sweep time for the unit field interval (taken in
dimensionless units), i.e. to define $t_{\rm red}$ as the quantity
which is reciprocal to the sweep rate averaged over the whole
intervals.

Figure 2 shows the dependences of the blocking temperatures defined
in different ways on $1/\ln (t_{\rm red})$: $T_{b}^{r}$ was
obtained from expression (\ref{eq13a}), while $T_{\rm
{b-scan}}^{r}$ was calculated numerically by means of the RM for
different values of $n$, where $n$ is the number of intervals,
into which the unit field interval $\Delta h=1$\ was divided. In
calculations, the $T_{\rm  {b-scan}}^{r}$ was regarded as a
value of the reduced temperature, at which the remanent magnetization was
$e$ times less than the saturation magnetization.

It is seen that all dependences, which are calculated with the use of the
RM, lie between two curves, one of which is the curve obtained from (\ref%
{eq13a}), while the other is some limiting curve, to which the
calculated results tend when $n$ goes to infinity. This curve
corresponds to the CSMF case. It is clear that the solution
corresponding to $n=1$, i.e. when one point falls on a unit
interval, coincides with that obtained from formula (\ref{eq13a}).
One can see that the results start to diverge from the limiting
solution already for $n\sim 200$.

The dependences $h_{c}(T_{\rm red})$ were also calculated for the case $%
n\rightarrow \infty $. Extrapolation of their low temperature regions to the
temperatures where $h_{c}\rightarrow 0$ resulted in the same dependences of $%
T_{\rm  {b-scan}}^{r}(t_{\rm red})$ as those which were
determined from the expression for the remanent magnetization. For
all $t_{\rm red}$
values, except for the shortest ones ($t_{\rm red}<5$), the curves $%
h_{c}(T_{\rm red})$ correspond to Eq. (\ref{eq18a}), in which $T_{\rm  {%
b-scan}}^{r}(t_{\rm red})$ substitutes for $T_{b\ast }^{r}$.

The analysis of the data in Fig. 2 shows that the CSMF case is the
limiting case of relaxation measurements when the measuring time
is reduced. In fact, in the former case, the system resides in the
saturated state for a considerable part of the sweep time which
falls on a unit field interval. The effective sweep time from the
saturating field to the zero~ (or~ coercive)

\begin{figure}
\includegraphics[width = 3.2 in]{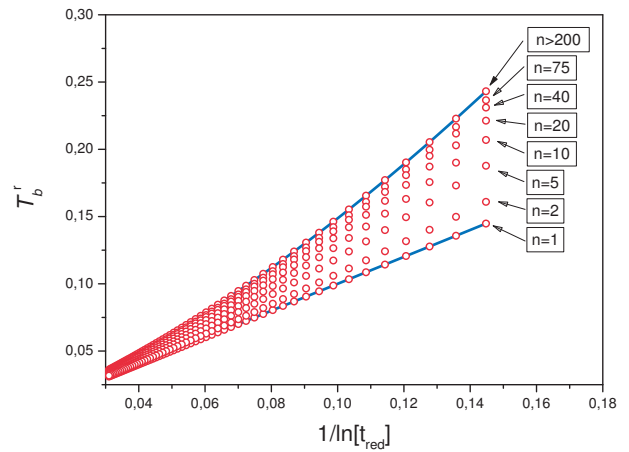}
\caption{\label{fig:epsart} The
$T_{\text{b-scan}}^{r}(t_{\rm red})$ dependences calculated by the
RM technique for various numbers of points per unit field interval
(circles). The top solid line corresponds to the limiting case
$n\rightarrow \infty $. The bottom line shows the
$T_{b}^{r}(t_{\rm red})$ dependence calculated by formula
(\ref{eq13a})}
\end{figure}

%
%

\noindent one turns out to be far shortest than $t_{\rm red}$
defined above for the CSMF protocol. For this protocol, one can
introduce the effective
measuring time $t_{\rm red}^{\rm  {eff}}$, at which $T_{\rm  {b-scan}%
}^{r}(t_{\rm red})$ in the CSMF case will coincide with $T_{b}^{r}(t_{\rm  {%
red}}^{\rm  {eff}})$ in the case of relaxation measurements, i.e.
it will equal $1/\ln (2t_{\rm red}^{\rm  {eff}})$. It is
appropriate to regard such time as the sweep period for the field,
at which the equilibrium magnetization changes from zero to a
certain value $aM_{0}$. Then, from the fitting of the limiting
curve in Fig. 2, we should find the optimal value of
$a$ (its expected value is about 2). Such an approach leads to the equation $t_{%
\rm  {red}}^{\rm  {eff}}=t_{\rm red}$arctg $h(a)/\left[ 2\ln (2t_{\rm  {%
red}}^{\rm  {eff}})\right] $, whose solution is%
\begin{equation}
t_{\rm red}^{\rm  {eff}}(t_{\rm red},a)={\frac{{t_{\rm red}}\, \rm  {%
arctanh }\,(a)}{2W\left[ {t_{\rm red}}\, \rm  {{arctanh
}}\,{(a)}\right] }}\rm  {.} \label{eq22a}
\end{equation}%
Here, $W(x)$ is the so-called Lambert $W$-function which is a
reciprocal function to $x=W(x)\exp (W(x))$. This function was introduced
into mathematical physics relatively recently [18]. In a number of
cases, for example in a popular program ``Mathematica'' developed
by ``Wolfram Research'' company, it is denoted as ``ProductLog''.
The examples of the solution of various tasks of mathematical
physics with the use of this function are presented in [19].

The calculations we carried out have shown that the expression%
\begin{equation}
T_{\rm  {b-scan}}^{r}(t_{\rm red})=1/\ln \left[ 2t_{\rm red}^{\rm  {eff%
}}(t_{\rm red},a=0.45167)\right]   \label{eq23a}
\end{equation}%
approximates the limiting curve of Fig. 2 with a high precision. Noticeable
deviations are observed only for the smallest  $t_{\rm red}$ values ($t_{%
\rm  {red}}<5$), which are of minor importance from the practical
point of view. Thus, this approximation requires only one fitting
parameter $a$.

Taking into account that the $W(x)$ function is not widely used in the
scientific literature, we also found the approximation of the
limiting curve for $T_{\rm  {b-scan}}^{r}(t_{\rm red})$ by a power
series of $\left[ \ln
(2t_{\rm red})\right] ^{-1}$:%
\[ T_{\rm b - scan}^{r} (t_{\rm red} ) = {\frac{{1}}{{\ln \,(2t_{\rm red}
)}}} \times\]
\begin{equation}
 \label{eq24a} \times\left( {1 + a_{0} + {\frac{{a_{1}}} {{\ln
\,(2t_{\rm red} )}}} + {\frac{{a_{2}^{}}} {{[\ln \,(2t_{\rm red}
)]^{2}}}} + {\frac{{a_{3}}} {{[\ln (2t_{\rm red} )]^{3}}}}.....}
\right)
\end{equation}
with four fitting parameters: $a_{0}=0.01$, $%
a_{1}=5.197$, $a_{2}=-3.581$, and $a_{3}=-1.602$. Finally, we
note that, in real CSMF experiments, ${t_{\rm red}}$ is usually
between $10^{10} $ and $10^{14}$. For this reason, the
discrepancies in the determination of a blocking temperature are
almost unnoticeable for different measurement methods.

\section{Recursion Method}
Before turning to a description of the calculation method for magnetic
reversal curves for different measuring times, it should be noted that
its implementation, as also in the case of MC modeling, requires some
efforts in the programming of a calculation procedure. We will use the same
dimensionless parameters as in Section 2: $h=Hm/(2K)$ is the magnetic field,
$t_{\rm red}=t\nu _{0}$ is the measuring time, $T_{\rm  {red}%
}=kT/(KV)$ is the temperature, $n$ is the number of points per
unit field interval, and $\overline{m_{\rm red}}=f\overline{m}/(fm_{s})=M_{\rm red}$- is average magnetization for measuring time, which normalized on a saturation magnetization $m_{s}$.

The method is based on the following prerequisites.

1. The hysteresis, which is a result of the metastability, will become apparent
at a finite measuring time only if the state of a system is
characterized by two minima in the dependence of its energy on the orientation
of a particle magnetic moment. The hysteresis originates from the
metastability with regard to thermally activated jumps over a potential
barrier. In its turn, the metastability appears as a result of the finiteness
of a measuring time.

2. In real magnetostatic measurements, the ratio of a barrier energy
at $H=0$ to a thermal energy, at which the deblocking of a magnetic
moment occurs, is near 25. That is why we assume that, even at
temperatures higher than the blocking temperature, the orientation
of a magnetic moment will be localized in one of the minima, rather
than smeared by temperature over a wide range of angles~$\theta $.

3. To determine the magnetization of a system, the concept of potential well
(minimum) populations is introduced. This concept is based on the distribution
statistics of magnetic moment directions in an infinitely great ensemble of
identical and equally oriented particles.

4. In dimensionless units, the magnetization of the system depends
only on the coordinates of minima ($\theta _{m1},\theta _{m2}$)
and their populations ($N_{1},N_{2}$): $M_{\rm red}=N_{1}\cos
(\theta _{m1})+N_{2}\cos (\theta _{m2})$.

The limits of applicability of this method will be discussed later on; we
will concentrate now only on the calculation procedure.

We reduce formula (\ref{eq2a}) for the density of energy of a separate particle
to that in the dimensionless units (the index $i$, which refers to the number of a
particle, will be omitted again, as was done in (\ref{eq11a})):%
\begin{equation}
U_{r}=-\cos ^{2}(\theta -\theta _{1})-2h\cos (\theta ).
\label{eq25a}
\end{equation}%
Consider the energy profile $U_{r}$\ in the phase space $\theta \in (-\pi ,\pi
]$ as a function of the magnetic field $h$. The energy minimum will correspond
to the equilibrium orientation of the magnetic moment. The number of extrema
can be found by solving the equation $\partial U_{r}/\partial
\theta =0$. Substituting the roots of this equation into the expression $%
\partial ^{2}U_{r}/\partial \theta ^{2}$, we can find out if a given root
corresponds to the energy minimum or maximum.

The method developed consists of a few successive  steps. At first, the
system is assumed to reside in a saturating magnetic field. Then we
sequentially change the field to smaller values and calculate the
magnetization, to which the system will come during the time interval which
is equal to the period of the system residence at a certain field point $%
h_{k}$. The relaxation time is determined by the form of the potential in a
field $h_{k}$. According to this method, the calculations should start from
a negative field, which is high enough so that the energy displays only one
minimum, and finish at a sufficiently high positive field, for which the
energy again displays only one minimum. Thus, the calculation procedure can
tentatively be divided into three stages.

The first stage is applicable when the field is negative and the state of
the system is characterized by only one energy minimum. This stage includes:

1. Determination of the total number of extrema in the range $\theta \in
(-\pi ,\pi ]$ and their separation into minima and maxima.

2. If the state of the system is characterized by only one energy minimum,
then $N_{1}=1$ and $N_{2}=0$, and the magnetization equals $M_{\rm  {red}%
}=N_{1}\cos (\theta _{m1})+N_{2}\cos (\theta _{m2})$. If there are two
minima, we should go on to the second stage.

3. Recording the magnetization for a given field point $h_{k}$, changing
the field to $h_{k+1}=h_{k}+\Delta $, and\ going on to item 1.

The second stage is applicable when the state of the system is characterized
by two energy minima and includes:

4. Determination of the total number of extrema  in the range $\theta \in
(-\pi ,\pi ]$ and their separation into minima and maxima.

5. If the state of the system is characterized by a single energy minimum,
then we should go on to the third stage. If there are two minima, a temporal
evolution of $N_{1}$ and $N_{2}$ should be considered. This includes:

a) determination of the equilibrium populations $N_{1\infty }$ and $%
N_{2\infty }$, i.e. the values of $N_{1}$ and $N_{2}$ when the
time interval is infinite:%
\begin{equation}
N_{1\infty }={\frac{{1}}{{1+\exp }\left[ {-(E}_{2}-E_{1}{)/T}_{\rm red}%
\right] }}\rm  {, \ \ \ }N_{2\infty }=1-N_{1\infty }\rm  {,}
\label{eq26a}
\end{equation}%
where $E_{1}=U_{r}(\theta _{m1})$ and $E_{2}=U_{r}(\theta _{m2})$ are the
energy values for the first and second minima, respectively;

b) tracing the relaxation of $N_{1}$ and $N_{2}$ for the time interval $%
t_{\rm  {loc}}=t_{\rm red}/n$:%
\begin{equation}
N_{1t}=N_{1\infty }+\left( {N_{1}-N_{1\infty }}\right) \,\exp \left( -t_{%
\rm  {loc}}/\tau _{\rm red}\right) \rm  {,}  \label{eq27a}
\end{equation}%
\begin{equation}
\tau _{\rm red}=\frac{1}{\exp \left[ -(E_{b}-E_{1})/T_{red}\right]
+\exp \left[ -(E_{b}-E_{2})/T_{red}\right] }\rm  {.}  \label{eq28a}
\end{equation}

c) recording the new values of $N_{1}$ and $N_{2}$:%
\[
N_{1}=N_{1t}\rm  {, \ }N_{2}=1-N_{1t}\rm  {.}
\]

6. Calculation of the magnetization $M_{\rm
red}=N_{1}\cos (\theta _{m1})+N_{2}\cos (\theta _{m2})$.

7. Fixation of the magnetization for a given field point $h_{l}$, changing
the field to $h_{l+1}=h_{l}+\Delta $, and going on to item 4.

The third stage is applicable when the field is positive, and there is no
second minimum. This stage includes:

8. Determination of the total number of extrema  in the range $\theta \in
(-\pi ,\pi ]$ and separation of them into minima and maxima.

9. If the state of the system is characterized by only one energy minimum,
then $N_{1}=0$ and $N_{2}=1$ and the magnetization equals $M_{\rm  {red}%
}=N_{1}\cos (\theta _{m1})+N_{2}\cos (\theta _{m2})$.

10. Fixation of the magnetization for a given field point $h_{m}$, changing
the field to $h_{m+1}=h_{m}+\Delta $, and going on to item 8.

The greater $n$, the more minutely the magnetization curve will be
calculated. At the same time, the real time interval spent on the
measurement of the complete hysteresis loop
(${\{}-h,h{\}},{\{}h,-h{\}}$) will equal $t=2t_{\rm red}\Delta
h/\nu _{0}$ s, where $\Delta h $ is the interval (taken in
dimensionless units) of the complete sweep of the magnetic field.
If one does not need to model the partial hysteresis loops, it is
enough to carry out the calculations for the interval
${\{}-h,h{\}}$, because the second interval ${\{}h,-h{\}}$ will be
symmetric relative to the coordinate origin ($0,0$). To obtain the
magnetization curves for an ensemble which is characterized by a
distribution of some particles' parameters, it is necessary to
divide the distribution function into sufficiently small intervals
and, having calculated the separate curves for each of the
intervals (i.e. for the average values of a parameter in this
interval), to sum them.\looseness=1

The inset in Fig. 3 shows the results of the modeling of magnetization curves
for the ensemble of uniaxial particles, whose easy axes are aligned in parallel
to the magnetic field. The calculations were made for $t_{\rm  {red}%
}=1.25\times 10^{6}$ and $n=125$. The temperature $T_{\rm red}$
was varied from 0 to 0.02. The temperature dependence of
coercivity (squares) is well described by formula (\ref{eq7a}) and
has a square-root character ($\alpha =2 $) for the given
orientation of the particle easy axes. The plot of this curve in
the form $h_{c}(\sqrt{T_{\rm red}})$ (triangles) confirms the
latter fact. Such a dependence is a straight line and its
extrapolation to the intersection with the ordinate axis
unambiguously determines the blocking temperature $T_{b}^{\rm
red}$= $(0.3\pm 0.005)^{2}$ = $0.09\pm 0.003$. At
the same time, the calculation of $T_{\rm  {b-scan}}^{r}(t_{\rm  {red}%
}=1.25\times 10^{6})$ according to formulas (\ref{eq23a}) and
(\ref{eq24a}) in the case $n\rightarrow \infty $ (or, at least,
$n\geq 200$) gives 0.0915. Thus, the value of $T_{b}^{\rm red}$ =
$0.09$ obtained from Fig. 3 well agrees with that found
from (\ref{eq23a}) and~(\ref{eq24a}).

The temperature dependence of the remanent mag\-ne\-ti\-za\-tion
$M_{r}$ normalized to the saturation magnetization exhibits a sharp
rise (see Fig. 3, circles) at temperatures where $h_{c}$ becomes
noticeable.

\begin{figure}
\includegraphics[width = 3.2 in]{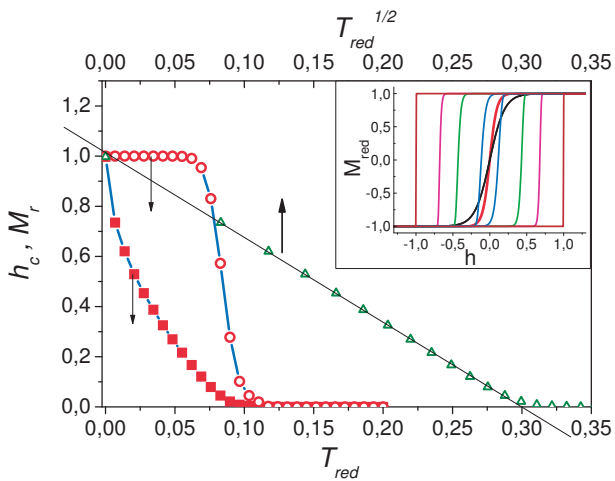}
\caption{\label{fig:epsart} Results of the modeling
of the $h_{c}(T_{\rm red})$ (solid
squares) and $M_{r}(T_{\rm red})$ (open circles) dependences for $t_{%
{\rm red}}=1.25\times 10^{6}$. The $h_{c}(\sqrt{T_{\rm red}})$
dependence is shown by open triangles. The calculations are
carried out for $n=125$. The inset shows the magnetization curves
which correspond to these dependences,
for $T_{\rm red}=0.2$, $0.1$ (anhysteretic curves) and $0.07$, $0.03$, $%
0.01$, and $0.00$ (with a hysteresis which consecutively widens)}
\end{figure}

%

\noindent

As was expected, the value $T_{\rm red}=$ $T_{b}^{\rm red}=0.09$
corresponds to a decrease in the remanence by a factor of 2.718. This
means that the plots of the temperature dependences of the coercivity
and the remanent
magnetization normalized to the blocking temperature (see expression (\ref%
{eq23a}) or (\ref{eq24a})) should coincide. Figure 4,{\it a} shows
the results of the RM modeling obtained on the same ensemble of
particles, but for different values of the measuring time. These
data replotted in
the coordinates where the abscissa axis is scaled by $T_{\rm  {b-scan}%
}^{r}(t_{\rm red})$ are presented in Fig. 4,{\it b}. It is seen
that
almost all the $H_{c}\left[ T_{\rm red}/T_{\rm  {b-scan}}^{r}(t_{\rm  {red%
}})\right] $ curves overlap one another. Replotting these curves
in the coordinates with a ``rooted abscissa'' shows that, at
temperatures far lower
than the blocking temperature, they are well described by formula (\ref{eq9a}%
).

A characteristic feature of the $M_{r}\left[ T_{\rm red}/T_{\rm  {b-scan}%
}^{r}(t_{\rm red})\right] $ dependence is a broadening of the front of the $%
M_{r}$\ growth with decrease in $t_{\rm red}$. All the curves
intersect one another at a point, whose ordinate equals 0.368, i.e.
$1/e$. Thus, the simulated
magnetization curves, as well as the $H_{c}(T_{\rm red})$ and $%
M_{r}(T_{\rm red})$ dependences obtained from them, are in
compliance with the regularities described in Section 2, which
determine their behavior over a wide range of measuring time
values.

The RM calculations of the magnetization curves were also carried out for
the ensembles of particles with 3D- and 2D-distributions of easy axes. The
results obtained well agree with the data calculated by the MC method [1,6,7].
The fact that the time of computer calculations is much less within the
RM than that within the MC method gives us an additional argument in favor of the
method we have proposed here.

It is worth noting that the method developed contains a series of
approximations, which can lead to some inaccuracy of the results
obtained. The most important approximation is related to a
failure to consider the thermal fluctuations of a magnetic
moment in the vicinity of an energy minimum . This, in turn, gives
rise to an inaccuracy in the determination of the magnetization at
sufficiently high temperatures (or for very short values of
the measuring time ($t_{\rm red}\leq 5$)). It should be stressed once
more that, in the first place, the RM is used for the calculations
of the \rm  it{hysteresis} loops of the magnetization. To study the
magnetization curves \rm  it{above} \rm  it{the blocking
temperature},\ it is enough to utilize formula (\ref{eq5a}). With
the use of this formula, one can also estimate the measure of
inaccuracy for the magnetization calculated by the RM at
temperatures higher than the blocking temperature. It is obvious
that the higher the temperature, the greater is the inaccuracy. To check
the role of this factor, we calculated the
magnetization curves for $t_{\rm red}=1.25\times 10^{4}$ and $T_{\rm  {red}%
}>$ $T_{b}^{\rm red}$ (namely, for $T_{\rm red}=0.2$) by the RM
and formula (\ref{eq6a}). The maximal error in the
determination of the magnetization did not exceed $1.5\%$. The
measuring times, which are shorter than the above value, are
hardly possible in practice.

\section{Monte--Carlo Method}

For the modeling by Monte--Carlo\ method, the standard algorithm
suggested by Metropolis et al. [16] was used. It is known that,
for a sufficiently great number of steps $N_{\rm  {MC}}$, such an
algorithm leads to the Boltzmann distribution. This means that a
system comes to the thermodynamic equilibrium and thus, no
metastability and, respectively, no hysteresis will be observed
unless we introduce some special tricks. In a general case, for a
great number of MC steps, the results will tend to those which can
be obtained with the use of formula (\ref{eq5a}). To ``catch'' the
metastability in the process of magnetization reversal, it is
necessary to use a finite number of MC steps and restrict the
generation of a trial random orientation of the magnetic moment in
the vicinity of the current orientation by a certain not great
aperture $\Delta \theta $, instead of the generation over the~
whole~ phase~ space.


\begin{figure}
\includegraphics[width = 3.2 in]{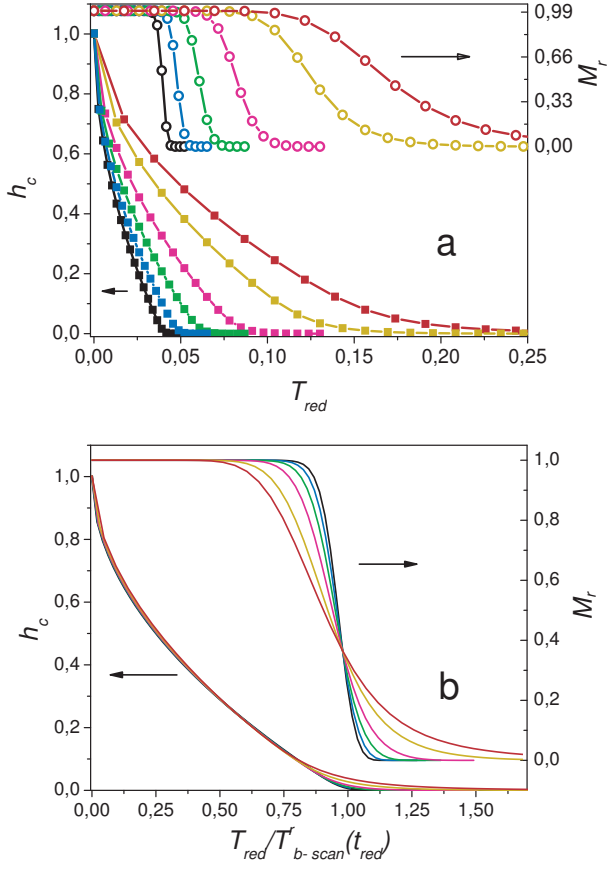}
\caption{\label{fig:epsart} (\textit{a}) Families of
the $H_{c}(T_{\rm red})$ (solid squares) and $M_{r}(T_{\rm red})$
(open circles) curves calculated for the ensemble of particles,
whose easy axes are aligned along the magnetic field,
for $t_{\rm red}=2.5\times 10^{12}$, $2.5\times 10^{10}$, $2.5\times 10^{8}$%
, $2.5\times 10^{6}$, $2.5\times 10^{4}$, and $2.5\times 10^{3}$
($n=250$). The
curves are shifted towards the right side as $t_{\rm red}$ decreases. (%
\textit{b}) The same curves replotted in the coordinates where the
abscissa axis is scaled by $T_{\rm  {b-scan}}^{r}(t_{\rm red})$}
\end{figure}

%


\noindent This trick is one of the standard MC techniques to model
the hysteresis loops [7, 10, 12]. The procedure of modeling is
divided into two stages: a thermalization of the system and a
magnetization reversal process itself.

The system thermalization is regarded as the pro\-ce\-dure
consisting from tens to hundreds of thousands of MC steps at high
negative fields and sufficiently high temperatures. This procedure
is aimed at bringing the system to a state which is equivalent to
the thermal equilibrium. To model the magnetization reversal
process, one should give an increment in the magnetic field and
perform a given number of MC steps for each field point. Such
algorithm is widely used, and we won't describe it in details (see
its description, for example, in [10,12,17]). Here, we only note
that, in our calculations, we
set the aperture value $\Delta \theta =6%
{{}^\circ}%
$ (the role of the aperture will be discussed below). The thermalization
procedure was carried out for $T_{\rm red}=0.5$, $h=-10$, and $N_{\rm  {MC%
}}=10000$. To compare the efficiency of the MC method and RM, we performed
a series of calculations of the magnetization curves for various $T_{\rm  {%
red}}$ by both methods. To determine the blocking temperature, we utilized the
extrapolation of a low-temperature region of the $h_{c}(\sqrt{T_{\rm red}}%
)$ dependence, which is linear in these coordinates, to its intersection
with the abscissa axis. In both cases, 300 field points  fell on one
magnetization curve ($-2<h<2$), which meant that $n$ was equal to $75$.
Then we carried out the calculations of the $H_{c}(T_{\rm red})$ and $%
M_{r}(T_{\rm red})$ dependences according to the MC method with $N_{\rm  {%
MC}}=5\times 10^{6}$, a measuring time $t_{\rm red}$\ for the RM
procedure was fitted in such a way that the dependences obtained
agreed as much as possible with the results of the MC
modeling.\looseness=1

Figure 5 presents the $H_{c}(T_{\rm red})$ and $M_{r}(T_{\rm
red})$ de\-pen\-den\-ces obtained by the MC method (squares and
circles) and RM (solid lines) for the ensemble of particles, whose
easy axes are aligned along the magnetic field. It is seen from
the figure that, in spite of the disadvantages of the RM, which
were formulated in Section 3, both the methods give almost
identical dependences. Since the duration of the MC calculations
was sufficiently long (it took 46 min to calculate one
magnetization curve), only a small number of particles (5
particles in the ensemble) was taken for calculations and this
resulted in a noticeable data scattering. On the contrary, it took
no more than a minute to make the RM calculations for one
magnetization curve. In the case under consideration, the number
of MC steps at each field point was $5\times 10^{6}$, while the RM
with a fitting procedure described above gave $t_{\rm  {loc}%
}=8\times 10^{3}$, which means that, for $n=75$, $t_{\rm red}=nt_{\rm  {loc}%
}=6\times 10^{5}$. Since the parameter $\nu _{0}$ is of the order of $%
10^{8}-10^{11}$, the actual measuring time for such a kind of the
MC procedure corresponds to $10^{-4}-10^{-7}$ s. This is, of
course, an extremely small time interval. At the~ same~ time,~ it~
is~ pertinent~ to~ specify~ the interrelation between the number
of MC steps and the aperture, on the one hand, and the~ real~
time~ interval,

\begin{figure}
\includegraphics[width = 3.2 in]{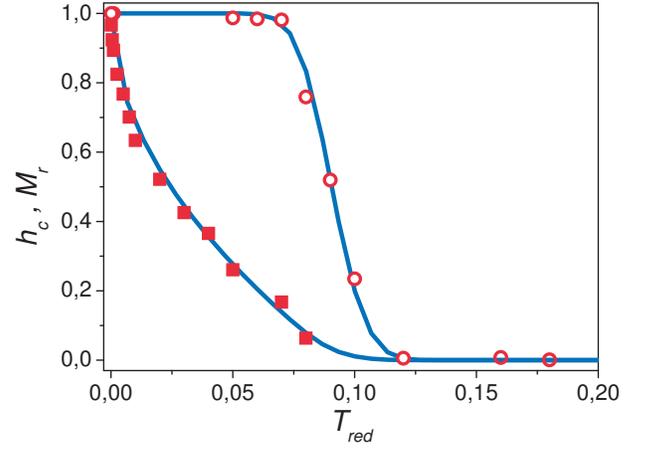}
\caption{\label{fig:epsart} Dependences $H_{c}(T_{\rm red})$ (solid squares) and $M_{r}(T_{\rm  {red}%
})$ (open circles) obtained by the MC modeling ($N_{\rm  {MC}%
}=5\times 10^{6}$, $t_{\rm  {loc}}=8\times 10^{3}$, $n=75$) and
the RM
(solid lines)}
\end{figure}

%
%

\section{Relationship Between the Parameters of MC Modeling and
Real Measuring Time}
It is appropriate to assume that, for the MC procedure, the
equivalent real measuring time $t_{\rm red}$ should be
proportional to $nN_{\rm  {MC}}$. For this reason, the results
obtained in Section 2 were used to search
for such a dependence. At first, the calculations of the blocking temperature $%
T_{b}^{\rm red}$ were carried out by both methods for different
measuring times (different numbers of MC steps).

Figure 6 shows the dependences obtained along with a theoretical
curve which corresponds to formulas (\ref{eq23a}) and
(\ref{eq24a}). To make a comparison of the results to be more
covenient, all the dependences were represented in the form
$T_{b}^{\rm red}\left[ \ln (t_{\rm  {sys}}/t_{\rm  {fit}})\right]
$, where $t_{\rm {sys}}$ is the effective measuring time
characteristic
of each method (for the RM and the theoretical dependence, $t_{\rm  {%
sys}}=t_{\rm red}$; for the MC procedure, $t_{\rm  {sys}}=nN_{\rm  {MC}}$%
), and $t_{\rm  {fit}}$ is the fitting parameter specific for each
procedure. It turned out that, for the RM procedure and the
theoretical dependence, $t_{\rm  {fit}}=1$. It was also confirmed
that, for the MC procedure, firstly, the equivalent measuring time
is actually proportional to $nN_{\rm  {MC}}$, and, secondly, the
coefficient of proportionality is about $450$. It is seen that the
points calculated by both methods agree well with the theoretical
curve. Based on this, the relation between the number of MC steps
and the equivalent measuring

\begin{figure}
\includegraphics[width = 3.2 in]{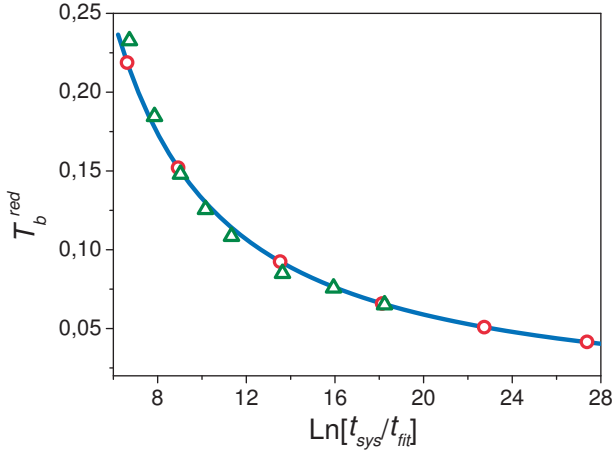}
\caption{\label{fig:epsart} Dependences of
$T_{b}^{\rm red}$ on the effective measuring time $t_{\rm  {sys}}$
(the number of MC steps) for the MC method (triangles) and the RM
(circles). The solid curve corresponds to the calculations by
formulas
(\ref{eq23a}) and (\ref{eq24a})}
\end{figure}

%

\noindent time was obtained in the form%
\begin{equation}
t_{\rm red}^{\rm  {MC}}\approx nN_{\rm  {MC}}/450.
 \label{eq29a}
\end{equation}

One can estimate the number of MC steps which corresponds
to the measuring time $t\nu _{0}=10^{10}$. It equals approximately
$5\times 10^{12}$.\ Taking into account that one MC step contains
a few tens of operations and that a modeling is performed for an
ensemble of particles, the necessary computational resources
exceed $10^{14}$ operations. Thus, it is concluded that
the computational capability of modern computers is not sufficient
to carry out the real time modeling. It is obvious that, for some
calculations, it is more important to deal with a great number of
particles in an ensemble and less important whether the number of
MC steps is great or small. However, to simulate, for example, the
ZFC/FC (zero field cooling/field cooling) procedure, the value of
the measuring time is likely to be decisive.

Let us discuss the origin of the parameter $t_{\rm  {fit}}$ for
the MC procedure. The only parameter which we have not yet varied
is the generation aperture $\Delta \theta $. It was noted that the
MC calculations above were
carried out under the condition that $\Delta \theta =6%
{{}^\circ}%
$. Let us start from the assumption that it is the parameter
$\Delta \theta $ that determines the $t_{\rm  {fit}}$ value. To
make sure of this, we performed the MC calculations of $T_{b}^{\rm
red}(N_{\rm  {MC}})$
dependences for various values of aperture ($\Delta \theta =1.5\div 48%
{{}^\circ}%
$).

\begin{figure}
\includegraphics[width = 3.2 in]{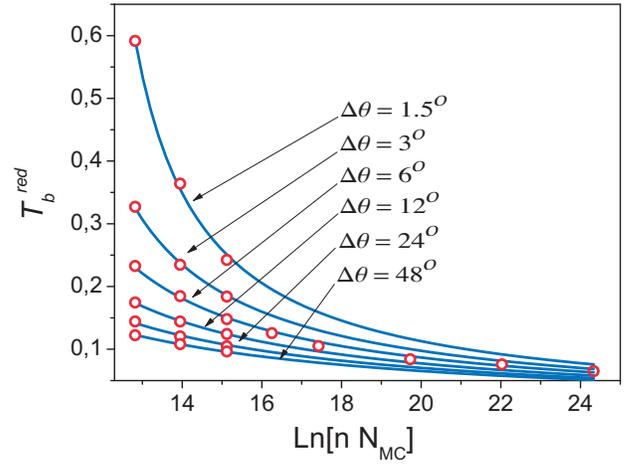}
\caption{\label{fig:epsart} The $T_{b}^{\rm
red}(nM_{\rm  {MC}})$ dependences calculated by
MC method for various values of aperture $\Delta \theta =1.5$, $3$, $6$, $12$%
, $24$, and $48^{\circ} $ (circles). Solid lines are the fitting
of the obtained results with the function $T_{\rm
{b-scan}}(nN_{\rm
{MC}}/t_{\rm  {fit}})$}
\end{figure}

Figure 7 presents the results of this modeling. As is seen from the
figure, for the calculations with a small number of MC steps, $T_{b}^{\rm  {%
red}}$ strongly depends on $N_{\rm  {MC}}$. As $N_{\rm  {MC}}$
increases, the aperture value to a lesser extent affects
$T_{b}^{\rm red}$. At the same time, a family of $T_{b}^{\rm
red}\left[ \ln (nN_{\rm  {MC}})\right]
$ dependences can be well approximated by a function $T_{\rm  {b-scan}}(nN_{%
\rm  {MC}}/t_{\rm  {fit}})$. It follows from general considerations that $t_{%
\rm  {fit}}(\Delta \theta )$ should be dependent on $U_{r}$, its
derivatives, or its integrals. However, we have not succeed in finding
the precise analytic
dependence. We can only note that, for $\Delta \theta <45%
{{}^\circ}%
$, the admissible  expression resulted from the fitting of the data in Fig.
7 is%
\begin{equation}
t_{\rm  {fit}}\approx 4.9/\left[ \sin ^{2}(\Delta \theta )\right]
. \label{eq30a}
\end{equation}%
Thus, we have obtained the final expressions which answer a question about which
measuring time corresponds to the MC calculations:%
\begin{equation}
t_{\rm red}^{\rm  {MC}}\approx \;{\frac{{\sin ^{2}(\Delta \theta )}}{{%
4.9\;}}}nN_{\rm  {MC}}, \label{eq31a}
\end{equation}%
\begin{equation}
~T_{b}^{\rm  {MC}}\approx T_{\rm  {b-scan}}\left[ n{\frac{{N_{\rm  {MC}}}}{{%
4.9}}}\sin ^{2}(\Delta \theta )\right] .\label{eq32a}
\end{equation}

\section{Conclusions}

In this work, the recursion method has been developed for the calculations
of the magnetic properties of the ensemble of single-domain particles. Its
applicability to the system of oriented particles with a uniaxial anisotropy
is demonstrated. There are no hindrances to apply such a procedure to the case
of a cubic anisotropy, introduce the distribution function for a certain
parameter, make the anisotropy constant temperature-dependent, or even
introduce the dipole-dipole interaction between the particles of an ensemble.
It is also not difficult to model the ZFC/FC procedure. In our opinion, the
RM results will far better reflect the real experiments, than the
results of the MC modeling.

The relation, which correlates the magnetic parameters of the ensemble
of uniaxially anisotropic magnetic particles with a measuring time for
these properties in various experimental procedures, is obtained.

It is shown that, depending on a kind of experiment, the relationship
between the blocking temperature and the measuring time has somewhat
different forms.

The calculations of the magnetization curves for the ensemble of
uniaxial single-domain particles are carried out for different
measuring times. The similar calculations performed by the
Monte-Carlo technique confirm the adequacy of the method developed
here. The latter method requires far less~ computational~
resources

%

\noindent in comparison with the modeling of an analogous task by
the Monte--Carlo method.

With the use of the new method, we succeeded in the establishment of the
empirical dependence between the number of MC steps and the generation aperture of
a random direction of the magnetic moment, on the one hand, and the measuring
time, which corresponds to these parameters, on the other hand. It is
shown that the parameters, which are usually used for the MC modeling of the
behavior of ensembles of magnetic particles, correspond to unlikely short
values of the measuring time.

\begin{flushright}
{\footnotesize Translated from Ukrainian by
A.I. Tovstolytkin}
\end{flushright}

\bibliography{}

[1] E.C. Stoner and E.P. Wohlfarth, Philos. Trans. Roy. Soc. London,
Ser. A  {\bf 240}, 599 (1948).\

[2] S.V. Vonsovskii, {\it Magnetism} (Moscow, Nauka, 1971), Chapter
23, Section 6 (in Russian).\

[3] L. Neel, Ann. Geophys. {\bf 5}, 99 (1949).\

[4] W.F. Brown, Phys. Rev. {\bf 130}, 1677 (1963).\

[5] H. Pfeiffer, Phys. Status Solidi A {\bf 118}, 295 (1990).\

[6] Lin He and Chinping Chen, Phys. Rev. B {\bf 75}, 184424 (2007).\

[7] J.~Garc\'{\i}a-Otero, M.~Porto, J.~Rivas, and A.~Bunde, J.~Appl.~Phys.~%
\textbf{85}, 2287 (1999).\

[8] O. Iglesias, A. Labarta, Physica B \textbf{372}, 247 (2005).\

[9] O. Iglesias, A. Labarta, Phys. Rev. B \textbf{63}, 184416
(2001).\

[10]D.A. Dimitrov and G.M. Wysin, Phys. Rev. B \textbf{54}, 9237
(1996).\

[11] L. Wang, J. Ding, H.Z. Kong, Y. Li, and Y.P. Feng, Phys. Rev. B \textbf{%
64}, 214410 (2001).\

[12] O.V. Billoni, S.A. Cannas, and F.A. Tamarit, Phys. Rev. B
\textbf{72}, 104407 (2005).\

[13] R.W. Chantrell, N. Walmsley, J. Gore, and M. Maylin, Phys. Rev. B
\textbf{63}, 024410 (2000).\

[14] M.A. Chuev, Pis'ma v Zh. Eksp. Teor. Fiz. \textbf{85}, 744 (2007).\

[15] O. Michele, J. Hesse, and H. Bremers, J. Phys.: Cond. Matter
\textbf{18}, 4921 (2006).\

[16] N. Metropolis, A.W. Rosenbluth, M.N. Rosenbluth, A.H. Teller, and
E. Teller, J. Chem. Phys. \textbf{21}, 1087 (1953).\

[17] C.M.P. Russell and K.M. Unruh, J. Appl. Phys. \textbf{99}, 08H909
(2006).\

[18] R.M. Corless, G.H. Gonnet, D.E.J. Hare {\it et al.}, Adv. Comput.
Math. \textbf{5}, 329 (1996).\

[19] S.R. Valluri D.J. Jeffrey, and R.M. Corless, Canadian J. Phys. \textbf{%
78}, 823 (2000).

\end{document}